\documentclass{article}
\usepackage{sw20nhj}
\usepackage{amsmath}
\usepackage{graphicx}
\usepackage{amsfonts}
\usepackage{amssymb}

\begin{document}

\author{A. A. Darmohval, V. K. Petrov, G. M. Zinovjev}
\title{{\Large Potential between adjoint sources in arbitrary representations}}
\date{\textit{\ 252143 Kiev, Ukraine. 10.7.1998}}
\date{\textit{\ 252143 Kiev, Ukraine. 10.7.1998}}
\date{\textit{\ 252143 Kiev, Ukraine. 10.7.1998}}
\maketitle
\begin{abstract}
The potential between sources in arbitrary representations of the gauge group
is studied on an anisotropic lattice in a spherical model approximation. It is
shown analytically that for half-integer $j$ and $j^{\prime}$ in the
confinement phase the potential rises linearly , whereas for integer $j$
$\ $and half-integer $j^{\prime}$ it rises$\ \ $infinitely which means a
strong suppression of the combination of such states . For integer $j$ and
$j^{\prime}$ the potential shows Debay screening and Coulomb behavior in the
deconfinement phase . It is also shown, that $\left\langle \chi^{\left(
j\right)  }\right\rangle \backsim\left\langle \chi\right\rangle ^{2j}$ when
$\left\langle \chi\right\rangle \gtrsim1$ and is in agreement with the mean
field theory prediction, and $\left\langle \chi^{\left(  j\right)
}\right\rangle \backsim\left\langle \chi\right\rangle $ for $\left\langle
\chi\right\rangle \lesssim1$ which agrees with MC experiment. String tension
model-computed for sources invariant under center group transformations
demonstrates Casimir scaling in the intermediate distance regime and turns
into zero at large distances.
\end{abstract}

%
%
\vfill\eject\pagestyle{plain}\setcounter{page}{1}

\section{Introduction.}

Statistical QCD evaluated on the lattice by means of computer simulation is
perhaps the only case in statistical physics where the critical behavior can
be calculated from the first principles of dynamics \cite{Creutz,McLSv,KPS}.
The power of the approach is best seen in the thermodynamics of pure $SU(N)$
gauge theory, i.e., for the systems consisting of gluons only
\cite{Creutz,B-P,Eng}. The main features considered are the deconfinement
transition and the properties of the hot gluon plasma \cite{Satz}. Lattice
gauge theory (LGT) provides a powerful tool to compute nonperturbative
properties of strong interactions, such as a potential between pairs of
infinitely heavy sources.

Its known from Monte Carlo simulations that probes in representations
insensitive to the centre gauge group transformation $Z(N)$ yield a screened
potential, while those sensitive to this transformations provide a linearly
rising one $F\approx\alpha R$ (where $\alpha$ - string tension, $R$ - distance
between sources) \cite{Mack,Pol,Suss}. As indicated in \cite{Damg-H}, the
potential between sources transforming as arbitrary irreducible
representations of the gauge group provides a good laboratory for testing the
dynamics behind screening and confining mechanisms. In particular, one can
measure the relevant distance scales, i.e. those that separate ``short
distances'' (essentially perturbative physics), from the ``long distances''.
The dynamics of the intermediate region has been found quite rich in many
aspects. Indeed, some very simple numerical simulations have shown that
adjoint sources, for example, may feel a linearly rising potential at
intermediate distances \cite{Cas1,Cas2,TW,Greensite,Bern-Amb}. Such sources
appear to ``deconfine'' at precisely the same critical temperature $T_{c}$\ at
which fundamental sources deconfine \cite{Damg87,FMM}. The string tension is
representation-dependent and appears to be roughly proportional to the
eigenvalue of the quadratic Casimir operator \cite{Cas1,Cas2,Cas3,TW}.

As the distance between color sources with zero ''n-ality'' increases they are
eventually screened by gluons. This asymptotic regime extends from the
color-screening length to infinity, and in the case of $SU(2)$\ gauge group
the string tensions become equal for half-integer $j$ -s\ and turn into zero
for integer ones. In particular, the string between quarks in an adjoint
representation must break at some distance that presumably depends on the mass
of ''gluelumps'' (i.e. on the energy of a gluon bound to a massive adjoint
quark) \cite{Poul-Trot}.

>From universality hypothesis it follows, in particular, that all higher
representations sensitive to the center of the gauge group should be
\textit{equivalent} order parameters with the same critical behavior as the
fundamental representation. However different behavior was found for the
higher representations at $SU(2)$\ lattice gauge theory in $\left(
3+1\right)  $\ dimensions \cite{Damg87,FMM}.

The purpose of this work is an analytical investigation of the potential
between sources in arbitrary representations of the gauge group near the
critical point on an anisotropic lattice in a spherical model approximation.

The application of spherical model in statistical physics has a long history
since the time it had been introduced to investigate critical phenomena in the
ferromagnet \cite{Kac} and until recently (see, e.g., \cite{Fra-Hen,Cap-Col}).
Stanley \cite{Stan} established the correspondence between the spherical and
the Heisenberg models.

Although this model is of no direct experimental relevance, it may provide
useful insight since many physical quantities of interest can be exactly
evaluated with its help. In this context, the spherical model is quite a
useful tool in providing explicit checks of general concepts in critical
phenomena, see \cite{Barb,Sing,Coni,Khor}. Recently \cite{Fra-Hen} it was
successfully used for studying the transitions between a paramagnetic, a
ferromagnetic and an ordered incommensurate phase (Lifshitz point). Models of
this kind were investigated extensively (see recent review in \cite{revselke}%
). The spherical model predicts reasonable values for critical exponents.
Moreover, a 'basic' set of exponent relations is also satisfied by spherical
model for $2<d<5$ \cite{Joyce}.

\section{LGT\ on anisotropic lattice.}

The simplest way to increase the accuracy in numerical study of LGT at large
distances and high temperatures is to use anisotropic lattices in which the
temporal spacing is much smaller than that in the spatial directions
($\xi=a_{\sigma}/a_{\tau}\gg1$, the Hamiltonian limit). This approach becomes
a very popular technique at the present time since it allows Monte Carlo
calculations to be carried out with reasonable computational resources and
reap many of the benefits of fine lattices, while still using cheap coarse
spatial ones. There are convincing arguments \cite{K82,B-S} that in the
anisotropic Wilson gauge action~ no coefficients are to be tuned in order to
restore space-time exchange symmetry up to O($a^{2}$) errors. On such lattices
heavy quarks will not suffer large lattice artifacts as long as their masses
are small in the units of $a_{\tau}$.

To compute the partition function on a lattice of a size $N_{\tau}\times
N_{\sigma}^{3}$ ($N_{\tau}$ is the temporal extent, $N_{\sigma}^{3}$ is the
spatial extent of a lattice) after \cite{SvY} (see also \cite{billo,aver}) we
use the anisotropic lattice
\begin{equation}
S_{G}=\beta_{\tau}S_{E}+\beta_{\sigma}S_{M},
\end{equation}
where $S_{E}$ is electric and $S_{M}$ magnetic part of the gluodynamics action
and
\begin{equation}
\beta_{\tau}\equiv\beta\widetilde{\xi}\left(  \xi,\beta\right)  \gg
\beta_{\sigma}\equiv\beta/\widetilde{\xi}\left(  \xi,\beta\right)
,\qquad\beta\equiv\frac{2N}{g^{2}},
\end{equation}
with%

\begin{equation}
\xi=a_{\sigma}/a_{\tau}.
\end{equation}

In the weak coupling region $\beta_{\tau,\sigma}/\beta\approx1+O\left(
1/\beta\right)  $ \cite{hasen1,shig} and in this case $\widetilde{\xi}$ does
not differ essentially from $\xi$. Moreover, in the ``naive'' limit
\begin{equation}
U_{\nu}\left(  x\right)  \simeq1+ia_{\nu}g_{\nu}A_{\nu}^{c}\left(  x\right)
T_{c} \label{4}%
\end{equation}
and one must put exactly $\widetilde{\xi}=\xi$. However, in LGT the effective
values of the integration variables $U_{\nu}\left(  x\right)  $ are far enough
from the ``naive'' limit (especially in the deconfinement region). The
deviation of $\widetilde{\xi}\ $from $\xi$ is regarded as 'quantum
corrections' or 'renormalisation'. The behavior of $\widetilde{\xi}\left(
\xi,\beta\right)  $ in the vicinity $\xi\sim1$ has been studied both
perturbatively \cite{kar82} and by non-perturbative methods
\cite{bkns88,s98,k98,engels,boyd,eik98}. Since we cannot exclude, that in the
area of the asymptotically large $\xi$ the behavior of $\widetilde{\xi}$ may
be essentially different, we will not specify the dependence of $\widetilde
{\xi}$ on $\xi$, but only assume that $\widetilde{\xi}\left(  \xi
,\beta\right)  $ may be made arbitrary large by increasing $\xi$ at fixed
$\beta$.

It is well known \cite{G-K,O} (see a review in \cite{Ukawa}) that there is a
variety of ways to get (in accordance with a universality hypothesis
\cite{SvY}) the effective action
\begin{equation}
-S_{eff}=\gamma\left(  \beta,\xi,N_{\tau}\right)  \operatorname{Re}\sum
_{l=1}^{3}\sum_{x}^{N_{\sigma}^{3}}\chi_{x}\chi_{x+l}^{\ast}, \label{seff}%
\end{equation}
expressed in terms of the Polyakov lines $\chi_{x}^{\left(  j\right)
}=\mathrm{Tr}\prod_{t=1}^{N_{\tau}}U_{0}^{\left(  j\right)  }\left(
x,t\right)  $ in $j-$representation of $SU(N)$ group (we omit index $j$ for
fundamental representation). The crucial assumption for obtaining $\left(
\ref{seff}\right)  $is to discard the magnetic part ($S_{M}\sim1/\widetilde
{\xi}$) of action. Then $S_{G}\left(  U\right)  \simeq S_{E}$ with
\begin{equation}
S_{E}=-\beta\widetilde{\xi}\sum_{x}\sum_{\nu=1}^{3}\operatorname{Re}%
\mathrm{Tr}\left\{  U_{0}\left(  x\right)  U_{\nu}\left(  x+0\right)
U_{0}^{\dagger}\left(  x+\nu\right)  U_{\nu}^{\dagger}\left(  x\right)
-1\right\}  . \label{form-8}%
\end{equation}

As a rule it can be done in a strong coupling approximation \cite{G-K,O}.
However, as it is pointed out in \cite{SvY} at high 'temperatures' $T_{SY}%
=\xi/\left(  a_{\tau}N_{\tau}\right)  \equiv\xi T$ and for the couplings one
gets $\beta_{\sigma}\backsim1/T_{SY}$ and $\beta_{\tau}\backsim T_{SY}$.Hence
we may assume that for finite $T$ the magnetic part of action may be neglected
\cite{aver} compared to the electric one ($S_{E}\sim\xi$) in a Hamiltonian
limit $\left(  \xi\gg1\right)  $ even at low temperatures $T$. Such an
assumption doesn't look as harmless as in a strong coupling case. Indeed,
there are serious reasons to believe that the magnetic part of the action may
be of crucial importance for creating the confining forces \cite{mack,yaffe,T}%
, even at high temperatures \cite{sim}. Therefore, strictly speaking, $QCD$
without a magnetic part may be considered in a weak coupling area only as a
specific ('toy')\textit{\ }model.

Summarizing for the partition function with static sources $\eta_{x}$ we may
eventually write%

\begin{equation}
Z\sim\int\exp\left\{  \operatorname{Re}\left(  \gamma\sum_{x,l}\chi_{x}%
\chi_{x+l}^{\ast}+\sum_{x}\eta_{x}\cdot\chi_{x}^{\ast}\right)  .\right\}
\prod_{x}^{N_{\sigma}^{3}}d\mu_{x}. \label{su2-1}%
\end{equation}

\section{$SU(2)$ gauge model}

In a simple case of $SU(2)$ gauge group one may write
\begin{equation}
d\mu_{x}=\left.  \sin^{2}\left(  \frac{\varphi_{x}}{2}\right)  \frac
{d\varphi_{x}}{2\pi}\right|  _{-2\pi}^{2\pi};\qquad\chi_{x}=\chi_{x}^{\ast
}=2\cos\frac{\varphi_{x}}{2},
\end{equation}
so by introducing new variables: $\sigma_{x}=2\cos\frac{\varphi_{x}}{2}$ and
$\widetilde{\sigma}_{x}=2\sin\frac{\varphi_{x}}{2}$ the partition function
$\left(  \ref{su2-1}\right)  $ can be rewritten as
\begin{equation}
Z\sim\int_{-\infty}^{+\infty}\exp\left\{  \gamma\sum_{x,l}\sigma_{x}%
\sigma_{x+l}+\sum_{x}\eta_{x}\sigma_{x}\right\}  \prod_{x}\delta\left(
\sigma_{x}^{2}+\widetilde{\sigma}_{x}^{2}-4\right)  d\sigma_{x}\widetilde
{\sigma}_{x}^{2}d\widetilde{\sigma}_{x}. \label{su2-2}%
\end{equation}

To compute the partition function we use the spherical model
approximation\footnote{We want to stress, that the spherical model
approximation does not break the global $Z\left(  2\right)  $ - invariance of
the model.}:
\begin{align}
\prod_{x}\delta\left(  4-\sigma_{x}^{2}+\widetilde{\sigma}_{x}^{2}\right)   &
\rightarrow\delta\left(  4N-\sum_{x}\left(  \sigma_{x}^{2}+\widetilde{\sigma
}_{x}^{2}\right)  \right) \label{su2-3}\\
&  =\int_{c-i\infty}^{c+i\infty}\frac{ds}{2\pi i}e^{s\left(  4N-\sum
_{x}\left(  \sigma_{x}^{2}+\widetilde{\sigma}_{x}^{2}\right)  \right)
},\nonumber
\end{align}
where constant $c$ is chosen so that the integration path is placed at the
right side of all singularities of the integrand to ensure the legitimacy of
interchanging the integration order over $ds$ and $d\sigma_{x}d\widetilde
{\sigma}_{x}$. Therefore we can rewrite $(\ref{su2-2})$ as
\begin{equation}
Z=const^{^{\prime}}\int dse^{4sN}\int_{-\infty}^{+\infty}\exp\left(
-\sigma_{x}A_{xx^{\prime}}\sigma_{x^{\prime}}+\eta_{x}\sigma_{x}\right)
d\sigma_{x}\int_{-\infty}^{+\infty}e^{-s\widetilde{\sigma}_{x}^{2}}%
\widetilde{\sigma}_{x}^{2}d\widetilde{\sigma}_{x}, \label{su2-4}%
\end{equation}
where
\begin{equation}
A_{xx^{\prime}}=s\delta_{x,x^{\prime}}-\gamma\sum_{l=1}^{d}\delta
_{x+l,x^{\prime}}.
\end{equation}

It is clear that integration over $\widetilde{\sigma}_{x}$ can be done
trivially
\begin{equation}
\int\prod_{x}^{N_{\sigma}^{3}}e^{-s\widetilde{\sigma}_{x}^{2}}\tilde{\sigma
}_{x}^{2}d\widetilde{\sigma}_{x}=\exp\left\{  -N_{\sigma}^{3}\frac{3}{2}%
\ln\left(  s-3\gamma\right)  +const\right\}  .
\end{equation}

Integration over $\sigma_{x}$ may be fulfilled after Fourier
transformation\footnote{Discrete variables $q_{l}$ for $N_{\sigma}\gg1$can be
considered as continuous: $0\leq q_{l}<2\pi.$}
\begin{equation}
\sigma_{x}=\sum_{k}\zeta_{k}e^{iqx},q_{l}\equiv\frac{2\pi k_{l}}{N_{\sigma}%
};\quad l=1;2;3;\quad k_{l}=0,...,N_{\sigma}-1
\end{equation}
which diagonalizes bilinear form
\begin{equation}
\sum_{x,x^{\prime}}\sigma_{x}A_{xx^{\prime}}\sigma_{x^{\prime}}=N_{\sigma}%
^{3}\sum_{k}\left|  \zeta_{k}\right|  ^{2}A\left(  q\right)  ,
\end{equation}
where
\begin{equation}
A\left(  q\right)  =s-\gamma\sum_{l=1}^{d}\cos q_{l}. \label{A}%
\end{equation}

Now we can do the integration over $\zeta_{k}$ and write for partition
function
\begin{equation}
Z=const\int_{c-i\infty}^{c+i\infty}ds\exp\left(  \sum_{x,x^{\prime}}\eta
_{x}A_{xx^{\prime}}^{-1}\eta_{x^{\prime}}+N_{\sigma}^{3}\left(  4s-b\left(
s\right)  -\frac{3}{2}\ln s\right)  \right)  ,
\end{equation}
where
\begin{equation}
b\left(  s\right)  =\frac{1}{2}\mathrm{Sp}\left\{  \ln A\right\}  =\frac{1}%
{2}\int_{0}^{2\pi}\left(  \frac{dq}{2\pi}\right)  ^{d}\ln\left(  s-\gamma
\sum_{l=1}^{d}\cos q_{l}\right)  ,\text{ }%
\end{equation}
and
\begin{equation}
3\gamma A_{xy}^{-1}=\int_{0}^{2\pi}\frac{e^{-iq_{_{3}}(x-y)}\left(  \frac
{dq}{2\pi}\right)  ^{3}}{\frac{s}{3\gamma}-\frac{1}{3}\sum_{l=1}^{3}\cos
q_{l}}=R_{3}\left(  \frac{s}{3\gamma};x-y\right)  , \label{A-1}%
\end{equation}
with
\begin{equation}
R_{3}\left(  \frac{s}{3\gamma};x\right)  =\int_{0}^{\infty}e^{-t\frac
{s}{3\gamma}}I_{x}\left(  \frac{t}{3}\right)  I_{0}^{2}\left(  \frac{t}%
{3}\right)  dt. \label{R}%
\end{equation}
and $I_{n}\left(  x\right)  $ - modified Bessel function.

Taking into account that
\begin{equation}
\sum_{x}A_{xy}^{-1}=\sum_{y}A_{xy}^{-1}=\frac{1}{s-3\gamma}, \label{mg}%
\end{equation}
we obtain for the uniform sources $\eta_{x}\equiv\eta_{0}$%
\begin{equation}
Z\simeq const\int_{c-i\infty}^{c+i\infty}ds\exp\left\{  N\Phi\left(  s\right)
\right\}  , \label{su2-5}%
\end{equation}
with
\begin{equation}
\Phi\left(  s\right)  \equiv4s-b\left(  s\right)  -\frac{3}{2}\ln s+\frac
{\eta_{0}^{2}}{4\left(  s-\gamma d\right)  }. \label{ph}%
\end{equation}

To calculate ($\ref{su2-5}$) we use the steepest descent method that in the
limit $N_{\sigma}^{3}\rightarrow\infty$ gives an accuracy of order $O\left(
\frac{1}{N_{\sigma}^{3}}\right)  $. The saddle point $s_{0}$ is determined by
$\Phi^{\prime}\left(  s_{0}\right)  =0$ and considering that one may write
\cite{Joyce} for $\frac{s_{0}}{3\gamma}-1\equiv\varepsilon\simeq0$.
\begin{equation}
R_{3}\left(  \varepsilon+1;0\right)  \approx\frac{3}{2}-\frac{3}{\pi}%
\sqrt{\frac{3}{2}}\cdot\sqrt{\varepsilon}, \label{R3}%
\end{equation}
we get
\begin{equation}
\Phi^{\prime}=4-\frac{1}{2}\left(  \frac{1}{2\gamma}-\frac{\sqrt{\frac{3}{2}}%
}{\pi\gamma}\sqrt{\varepsilon}\right)  -\frac{1}{2\gamma\left(  1+\varepsilon
\right)  }-\frac{\eta_{0}^{2}}{4\left(  3\gamma\right)  ^{2}\varepsilon^{2}}.
\label{sad}%
\end{equation}
\textrm{\ }

Hence for $\gamma<\gamma_{c}$ \ the solution of $(\ref{sad})$ for $\eta=0$ is
given by
\begin{equation}
s_{0}\simeq3\gamma\left(  1+\frac{2}{3}\left(  8\pi\gamma\right)  ^{2}\left(
\frac{\gamma_{c}}{\gamma}-1\right)  ^{2}\theta\left(  \gamma_{c}%
-\gamma\right)  \right)  , \label{S0}%
\end{equation}
where the critical coupling $\gamma_{c}$ is defined as the solution of
$\Phi^{\prime}=0$ at the lowest value of $s_{0}\rightarrow3\gamma+0$ (and
$\eta\rightarrow0),$ which evidently gives $\gamma_{c}=3/16.$ In a vicinity of
the critical point one may write for $A_{00}^{-1}$
\begin{equation}
\frac{1}{2}A_{0,0}^{-1}\ \simeq\frac{1}{4\gamma}-\frac{3}{2}\left(  \frac
{1}{2\gamma}-\frac{1}{2\gamma_{c}}\right)  \theta\left(  \gamma_{c}%
-\gamma\right)  . \label{Ao}%
\end{equation}

It is easy to see, that in the deconfinement region $\left(  \gamma>\gamma
_{c}\right)  $ the solution of $\Phi^{\prime}=0$ with $\eta\equiv0$ is absent.
The only way to regain the solution in such region is to tend $s_{0}%
\rightarrow3\gamma$, because in this case the term $\frac{\eta_{0}^{2}%
}{\left(  s_{0}-3\gamma\right)  ^{2}}$ must be preserved even in a limit
$\eta_{0}\rightarrow0.$ Now the solution
\begin{equation}
s_{0}\simeq3\gamma+\frac{\eta/2}{\sqrt{1-\frac{\gamma_{c}}{\gamma}}%
}\rightarrow3\gamma
\end{equation}
sticks to the point $3\gamma$ so in this region we can write%

\begin{equation}
\frac{1}{2}A_{00}^{-1}\ \simeq\frac{1}{4\gamma}\left(  1-\frac{\sqrt{\eta
}\left(  1-\frac{\gamma_{c}}{\gamma}\right)  ^{-\frac{1}{4}}}{\sqrt{2\gamma
}\pi}\right)  \simeq\frac{1}{4\gamma},
\end{equation}
and
\begin{equation}
\left\langle \chi\right\rangle =\left[  \left(  \frac{1}{2}\eta_{x}%
A_{x,0}^{-1}\right)  _{s=s_{0}}\right]  _{\eta_{0}\rightarrow0}\simeq2\left(
1-\frac{\gamma_{c}}{\gamma}\right)  ^{\frac{1}{2}}\theta\left(  \gamma
-\gamma_{c}\right)  . \label{spin}%
\end{equation}
Hence in the deconfinement region $\left\langle \chi\right\rangle \neq0$ but
steadily approaches zero when $\gamma\rightarrow\gamma_{c}+0$ in accordance
with the well known fact, that $SU(2)$ - gluodynamics undergoes second order
phase transition.

For the two point correlation function for $R\gg1$ we get
\begin{equation}
\left\langle \chi_{0}\chi_{R}\right\rangle =\frac{A_{0R}^{-1}\left(
s_{0}\right)  }{2}+\left\langle \chi\right\rangle ^{2}\simeq c\frac{e^{-\alpha
R}}{R}+4\left(  1-\frac{\gamma_{c}}{\gamma}\right)  \theta\left(
1-\frac{\gamma_{c}}{\gamma}\right)  , \label{AR}%
\end{equation}
where$\ c=const$ and string tension (in lattice units) is given by
\begin{equation}
\alpha\simeq3\pi\left(  1-\frac{\gamma}{\gamma_{c}}\right)  \theta\left(
1-\frac{\gamma}{\gamma_{c}}\right)  . \label{tension}%
\end{equation}

The expression for correlation function $(\ref{AR})$ as $(\ref{R})$ is just
the Fourier transform of the propagator$~\left[  D(q)+m^{2}\right]  ^{-1}%
$where $D(q)\equiv\sum_{i=1}^{d}4\sin^{2}\frac{q_{i}}{2}~$\ is the lattice
laplacian in $d$--dimensional momentum space. In the perturbation theory
expression $(\ref{R})$ can be attributed to the lattice one-particle exchange.
A basic non-perturbative feature of the high temperature plasma phase of QCD
is the occurrence of the gluon screening mass \cite{HKR}, so parameter
$\alpha$, defined by $(\ref{tension})$ can be interpreted as screening mass as
well. The expression $(\ref{R})$ is not specific for the spherical model and
appears in many popular approaches, for example in the matrix model (see e.g.,
\cite{EMN}) and at the perturbation expansion ($g^{2}\ll1$) (see
\cite{CMV,CHMV}).

A more detailed case is considered in the Appendix. Here we present only the
final results
%
%
: \begin{table}[ptb]
\label{tab2}
%
%
\begin{tabular}
[c]{|l|l|}\hline\hline
\multicolumn{1}{||l|}{$j\text{ \&}j^{\prime}$} & \multicolumn{1}{||l||}{\qquad
\ \qquad\qquad\qquad\ $\left\langle \chi_{0}^{\left(  j\right)  }\chi
_{R}^{\left(  j^{\prime}\right)  }\right\rangle $}\\\hline\hline
$\text{both}\ \text{integer}$ & $\left\langle \chi^{\left(  j\right)
}\right\rangle \left\langle \chi^{\left(  j^{\prime}\right)  }\right\rangle
\exp\left\{  \left\langle \chi\right\rangle ^{2}A_{0,R}^{-1}c_{1}+\left(
j^{\prime}j-\frac{1}{4}\right)  \left(  A_{0,R}^{-1}\right)  ^{2}%
c_{2}\right\}  $\\\hline
$\text{both}\ \text{half-int.}$ & $\left\langle \chi^{\left(  j\right)
}\right\rangle \left\langle \chi^{\left(  j^{\prime}\right)  }\right\rangle
+q\left(  j^{\prime};j\right)  \frac{2A_{0,R}^{-1}}{A_{0,0}^{-1}}$\\\hline
$j\text{ int.};\text{ }j^{\prime}\text{half- int.}$ & $\left\langle
\chi^{\left(  j\right)  }\right\rangle \left\langle \chi^{\left(  j^{\prime
}\right)  }\right\rangle +\left\langle \chi^{\left(  j-\frac{1}{2}\right)
}\right\rangle \left\langle \chi^{\left(  j^{\prime}-\frac{1}{2}\right)
}\right\rangle A_{0,R}^{-1}c_{3}$\\\hline
\end{tabular}
\par
\vskip0.3cm \textbf{Tab. I}\textit{~~ }Pair correlation for states in higher
representations\end{table}We work in the parameter area $\gamma\sim\gamma_{c}$
and $R\gg1$ where$A_{0,R}^{-1}$ $/A_{0,0}^{-1}\ll1;$ and $\left\langle
\chi\right\rangle \ll1\ $so we preserve only the terms with lower powers of
$A_{0,R}^{-1}/A_{0,0}^{-1}$ $\ $and $\left\langle \chi\right\rangle $.

It is easy to see that in all cases the correlation function
\begin{equation}
C^{\left(  j\right)  \left(  j^{\prime}\right)  }\equiv\left\langle \chi
_{0}^{\left(  j\right)  }\chi_{R}^{\left(  j^{\prime}\right)  }\right\rangle
-\left\langle \chi^{\left(  j\right)  }\right\rangle \left\langle
\chi^{\left(  j^{\prime}\right)  }\right\rangle
\end{equation}
exponentially decreases in the confinement phase $\left(  \left\langle
\chi\right\rangle =0\right)  $ and can be used for the computation of
$\alpha,$ which plays a role of string tension when $\left\langle
\chi^{\left(  j\right)  }\right\rangle \left\langle \chi^{\left(  j^{\prime
}\right)  }\right\rangle =0$ and may be interpreted as a screening mass for
$\left\langle \chi^{\left(  j\right)  }\right\rangle \left\langle
\chi^{\left(  j^{\prime}\right)  }\right\rangle \neq0.$

\section{Conclusions.}

As it can be seen from TABLE 1, our model gives linearly rising free energy
$F\sim\alpha R$ in the confinement phase for half-integer $j$ and $j^{\prime}%
$. The potential between states with $j$ integer$\ $and $j^{\prime}$
half-integer (or vice versa) turns into infinity which means strong
suppression of such combination of states. For integer $j$ and $j^{\prime}$ we
get the screening potential: $F\sim-\frac{e^{-2\alpha R}}{R}$. In the
deconfinement phase $\left(  \left\langle \chi\right\rangle \neq0\right)  $ we
get $F\sim-1/R$. We would like to stress here that in the deconfinement phase
$\left(  \left\langle \chi\right\rangle \neq0\right)  $ for integer $j$ and
$j^{\prime}$ the screening mass is twice as much than in confinement $\left(
\left\langle \chi\right\rangle =0\right)  $ one.

As it was predicted by the mean field theory \cite{Damg}
\begin{equation}
\left\langle \chi^{\left(  j\right)  }\right\rangle \backsim\left\langle
\chi\right\rangle ^{2j}. \label{jj}%
\end{equation}
This result was confirmed in MC simulations \cite{Red-Satz} and as it can be
seen from TABLE 1 roughly agrees with such data when $\left\langle
\chi\right\rangle \gtrsim1$. However, as it was pointed out in \cite{Kis} in
the vicinity of critical point the relation $\left(  \ref{jj}\right)  $ does
not agree with MC experiment. TABLE also shows, that our result $\left(
\ref{vic}\right)  $
\begin{equation}
\left\langle \chi^{\left(  j\right)  }\right\rangle \backsim\left\langle
\chi\right\rangle ,
\end{equation}
obtained in such area agrees with the data given in \cite{Kis} and with
theoretical predictions developed both in framework flux picture in a strong
coupling approximation and within field theory ($\phi^{4}$) on the basis of
universality arguments \cite{Kis}.

Finally, there is the issue of the Casimir scaling (whose importance were
emphasized in \cite{Cas1,Cas2})\textrm{\ }of higher-representation string
tensions in the intermediate distance regime. For values $R$\ far from
asymptotic region $\left(  \alpha R\lesssim1\right)  $\ we can write
$A_{0,R}^{-1}\approx c_{0}\cdot\left(  1-\alpha R\right)  $\ and $\left\langle
\chi^{\left(  j\right)  }\chi^{\left(  j^{\prime}\right)  }\right\rangle
\simeq\left\langle \chi^{\left(  j\right)  }\right\rangle \left\langle
\chi^{\left(  j^{\prime}\right)  }\right\rangle \exp\left\{  -\alpha
\cdot\left(  j^{\prime}j-\frac{1}{4}\right)  cR+const\right\}  $\ where
$c=2c_{2}c_{0}^{2}$ \ is independent either from $R$\ or from $j$\ and
$j^{\prime}$\ so it follows from $\left(  \ref{casi1}\right)  $\ and $\left(
\ref{casi2}\right)  $:$\qquad$
\begin{equation}
\alpha\left(  j;j^{\prime}\right)  \backsim\alpha\cdot\left(  j^{\prime
}j-\frac{1}{4}\right)  .
\end{equation}
Therefore, the adjoint sources may 'feel' a linearly rising potential at
intermediate distances, and appear to ``deconfine'' at precisely the same
critical point at which the fundamental sources really deconfine. This
qualitatively agrees with Casimir scaling observed in such parameter region
\cite{Cas1,Cas2}. It is easy to see, that at very large distance scales
Casimir scaling breaks down and color screening is set.

As it was pointed long ago, sources trivially transformed under centre gauge
group transformations can not be used as an order parameter. An average value
of a pair of such sources also is not a good laboratory for critical phenomena
studies, because the magnetization $M^{\left(  j.j^{\prime}\right)  }%
=\lim_{R\rightarrow\infty}\left\langle \chi_{0}^{\left(  j\right)  }\chi
_{R}^{\left(  j^{\prime}\right)  }\right\rangle =\left\langle \chi^{\left(
j\right)  }\right\rangle \left\langle \chi^{\left(  j^{\prime}\right)
}\right\rangle $ differs from zero and masks the Debay term. Therefore
$C_{R}^{j,j^{\prime}}=\left\langle \chi_{0}^{\left(  j\right)  }\chi
_{R}^{\left(  j^{\prime}\right)  }\right\rangle -\left\langle \chi^{\left(
j\right)  }\right\rangle \left\langle \chi^{\left(  j^{\prime}\right)
}\right\rangle \rightarrow0$ with $R\rightarrow\infty$ so $\chi_{0}^{\left(
j\right)  }$ and $\chi_{R}^{\left(  j^{\prime}\right)  }$ are statistically
independent at asymptotically large distances and long range order is absent
either for $\left\langle \chi^{\left(  j\right)  }\right\rangle \left\langle
\chi^{\left(  j^{\prime}\right)  }\right\rangle =0$ or for $\left\langle
\chi^{\left(  j\right)  }\right\rangle \left\langle \chi^{\left(  j^{\prime
}\right)  }\right\rangle \neq0.$ On the other hand, the fact that the
correlation function $C_{R}^{j,j^{\prime}}$ exponentially decreases for all
$j$ and $j^{\prime}$ can be used for determination of string tension even in
the cases when $\left\langle \chi^{\left(  j\right)  }\right\rangle $ or
$\left\langle \chi^{\left(  j^{\prime}\right)  }\right\rangle $ are invariant
under $Z(N)$ transformations.

We also wish to note that computing the Binder cumulant $B$, defined as
\cite{Bind}:
\begin{equation}
B=\frac{1}{2}\left(  3-\frac{\langle\chi^{4}\rangle}{\langle\chi^{2}%
\rangle^{2}}\right)  ,
\end{equation}
we get for a case $SU(2)$ gauge group
\begin{equation}
B=\left(  1+\frac{2A_{00}^{-1}}{\left\langle \chi\right\rangle ^{2}}\right)
^{-2}.
\end{equation}
Therefore, in a vicinity of a critical point we get $B\approx0$ because
$\left\langle \chi\right\rangle =0$ in the confinement region and in
deconfinement region, if we are close enough to the critical point. Vanishing
of $B$ means that the distribution of $\chi$ is Gaussian, therefore the
spherical model must give reliable results in both regions.

\section{Appendix}

The $n$ - order correlations $\left\langle \prod_{k=1}^{n}\chi_{x_{k}%
}^{\left(  j_{k}\right)  }\right\rangle $ may be easily computed after
Fouirier transformation of $\chi^{\left(  j\right)  }=\mathrm{U}_{2j}\left(
\frac\chi2\right)  $
\begin{equation}
\mathrm{U}_{r}\left(  \frac\sigma2\right)  =\int_{-\infty}^{\infty}%
u_{r}\left(  z\right)  e^{i\sigma z}\frac{dz}{2\pi};\quad u_{r}\left(
z\right)  =\int_{-\infty}^{\infty}\mathrm{U}_{r}\left(  \frac\sigma2\right)
e^{-i\sigma z}d\sigma\label{FU}%
\end{equation}
and we get in a spherical model approximation%

\begin{align}
\left\langle \prod_{k=1}^{n}\chi_{x_{k}}^{\left(  j_{k}\right)  }%
\right\rangle  &  =\int_{-\infty}^{\infty}\left\langle \exp\left\{  \sum
_{x}\eta_{x}\sigma_{x}\right\}  \right\rangle \prod_{k=1}^{n}u_{2j_{k}}\left(
z_{k}\right)  \prod_{k=1}^{n}\frac{dz_{k}}{2\pi}\nonumber\\
&  =\int_{c-i\infty}^{c+i\infty}ds\int_{-\infty}^{\infty}\exp\left\{
-\sigma_{x}A_{x,x^{\prime}}\sigma_{x^{\prime}}+\eta_{x}\sigma_{x}\right\}
\prod_{k=1}^{n}u_{2j_{k}}\left(  z_{k}\right)  \frac{dz_{k}}{2\pi}\prod
_{x}^{{}}d\sigma_{x}%
\end{align}
with%

\begin{equation}
\eta_{x}=\eta_{0}+i\sum_{k=1}^{n}z_{k}\delta_{x}^{x_{k}};\quad\eta
_{0}\rightarrow0.
\end{equation}
Therefore after integration over $\sigma_{x}$
\begin{equation}
\left\langle \prod_{k=1}^{n}\chi^{\left(  j_{k}\right)  }\left(
\varphi\left(  x_{k}\right)  \right)  \right\rangle =\frac{\pi^{\frac
{N_{\sigma}^{3}}{2}}}{\sqrt{\det A}}\int_{-\infty}^{\infty}\prod_{k=1}%
^{n}u_{2j_{k}}\left(  z_{k}\right)  \exp\left\{  \frac{1}{4}\eta
_{x}A_{x,x^{\prime}}^{-1}\eta_{x^{\prime}}\right\}  \prod_{k=1}^{n}%
\frac{dz_{k}}{2\pi}%
\end{equation}
and taking into account $\left(  \ref{spin}\right)  $%

\begin{align*}
\lim_{\eta_{0}\rightarrow0}\frac{1}{4}\eta_{x}A_{x,x^{\prime}}^{-1}%
\eta_{x^{\prime}}  &  =-\frac{1}{4}z_{k}A_{x_{k},x_{n}}^{-1}z_{n}%
+iz_{k}\left[  \left(  \frac{1}{2}\eta_{x}A_{x,x_{k}}^{-1}\right)  _{s=s_{0}%
}\right]  _{\eta_{0}\rightarrow0}\\
&  =-\frac{1}{4}z_{k}\alpha_{k,r}^{-1}z_{r}+i\left\langle \chi\right\rangle
\sum_{k}z_{k},
\end{align*}
we obtain inverting Fourier transformation
\begin{equation}
\left\langle \prod_{k=1}^{n}\chi_{x_{k}}^{\left(  j_{k}\right)  }\right\rangle
=\overline{\prod_{k=1}^{n}\mathrm{U}_{2j_{k}}\left(  \frac{\sigma_{k}}%
{2}\right)  },
\end{equation}
with
\begin{equation}
\overline{\mathrm{Q}}\equiv\frac{\sqrt{\det\alpha}}{\pi^{\frac{n}{2}}}%
\int_{-\infty}^{\infty}\mathrm{Q}\exp\left\{  -\sum_{k,r=1}^{n}\left(
\sigma_{k}-\left\langle \chi\right\rangle \right)  \alpha_{k,r}\left(
\sigma_{r}-\left\langle \chi\right\rangle \right)  \right\}  \prod_{k=1}%
^{n}d\sigma_{k},
\end{equation}
where $\alpha_{k,r}$ is $n\times n$ matrix is defined by the condition
$\alpha_{k,r}^{-1}=A_{x_{k},x_{r}}^{-1}\left(  s_{0}\right)  $. In particular
one can write
\begin{align}
\left\langle \chi^{n}\right\rangle  &  =\left(  -D/2\right)  ^{\frac{n}{2}%
}H_{n}\left(  \left\langle \chi\right\rangle /\sqrt{-2D}\right)  \ \approx\\
&  \left\{
\begin{array}
[c]{cc}%
\begin{array}
[c]{cc}%
\left(  1+k\left\langle \chi\right\rangle ^{2}/D\right)  \left(  D/2\right)
^{\frac{n}{2}}n!/\left(  \frac{n}{2}\right)  ! & \left[  \frac{n}{2}\right]
=\frac{n}{2};\\
\left\langle \chi\right\rangle \left(  D/2\right)  ^{\frac{n-1}{2}}n!/\left(
\frac{n-1}{2}\right)  ! & \left[  \frac{n}{2}\right]  =\frac{n-1}{2};
\end{array}
& \gamma<2\gamma_{c},\\
\left\langle \chi\right\rangle ^{n}+\left\langle \chi\right\rangle
^{n-2}Dn\left(  n-1\right)  /4 & \gamma>2\gamma_{c},
\end{array}
\right. \nonumber
\end{align}
where the dispersion $D\equiv\left\langle \left(  \chi-\left\langle
\chi\right\rangle \right)  ^{2}\right\rangle $ is given by%

\begin{equation}
D=\left\langle \chi^{2}\right\rangle -\left\langle \chi\right\rangle
^{2}=A_{0,0}^{-1}/2=\left\{
\begin{array}
[c]{cc}%
1/4\gamma; & \gamma\gtrsim\gamma_{c},\\
4-\frac{1}{2\gamma}; & \frac{2}{3}\gamma_{c}\lesssim\gamma\lesssim\gamma_{c},
\end{array}
\right.  \leq\frac{4}{3}.
\end{equation}

The average value of the character $\chi^{\left(  j\right)  }$ in an arbitrary
irreducible representation $j$ may be computed as\footnote{Here $\left[
j\right]  $ refers to an integer part of $j$ and $H_{n}\left(  x\right)  $ are
Hermitian polinomials of order $n$.}%

\begin{equation}
\left\langle \chi^{\left(  j\right)  }\right\rangle =\sum_{m=0}^{\left[
j\right]  }\left(  -1\right)  ^{m}\binom{2j-m}{m}\left(  \sqrt{-D/2}\right)
^{2j-2m}H_{2j-2m}\left(  \left\langle \chi\right\rangle /\sqrt{-2D}\right)  ,
\end{equation}
or
\begin{equation}
\left\langle \chi^{\left(  j\right)  }\right\rangle =\sum_{m=0}^{\left[
j\right]  }\left(  -1\right)  ^{m}\binom{2j-m}{m}\left\langle \chi
^{2j-2m}\right\rangle , \label{j-U}%
\end{equation}
that for $\gamma\lesssim2\gamma_{c},$ meaning $\left\langle \chi\right\rangle
^{2}\lesssim2D$ immediately gives%

\begin{equation}
\left\langle \chi^{\left(  j\right)  }\right\rangle \simeq\left\{
\begin{array}
[c]{cc}%
\frac{3}{4}\frac{\left(  2j\right)  !}{j!}\left(  2\left(  1-\frac{1}{8\gamma
}\right)  \right)  ^{j}\left(  1+j\frac{\left\langle \chi\right\rangle ^{2}%
}{4\left(  1-\frac{1}{8\gamma}\right)  }\right)  ; & \left[  j\right]  =j,\\
\left\langle \chi\right\rangle \frac{\left(  2j\right)  !}{\left(  j-\frac
{1}{2}\right)  !}\left(  2\left(  1-\frac{1}{8\gamma}\right)  \right)
^{j-\frac{1}{2}}; & \left[  j\right]  =j-\frac{1}{2}.
\end{array}
\right.  \label{vic}%
\end{equation}

In the 'deep' deconfinement area $\gamma\gtrsim2\gamma_{c}$ or $\left(
\left\langle \chi\right\rangle ^{2}\gtrsim2D\right)  $%
\begin{equation}
\left\langle \chi^{\left(  j\right)  }\right\rangle \simeq\mathrm{U}%
_{2j}\left(  \frac{\left\langle \chi\right\rangle }{2}\right)  \simeq
\left\langle \chi\right\rangle ^{2j}\left(  1-\frac{j\left(  j+1\right)
}{\left\langle \chi\right\rangle ^{2}}\right)  . \label{j-hi}%
\end{equation}

The pair correlation%

\begin{align}
\left\langle \chi_{1}^{\left(  j_{1}\right)  }\chi_{2}^{\left(  j_{2}\right)
}\right\rangle  &  =\overline{\mathrm{U}_{2j_{1}}\left(  \frac{\sigma_{1}}%
{2}\right)  \mathrm{U}_{2j_{2}}\left(  \frac{\sigma_{2}}{2}\right)  }%
=\frac{\sqrt{\det\alpha}}{\pi}\int_{-\infty}^{\infty}\mathrm{U}_{2j_{1}%
}\left(  \frac{\sigma_{1}}{2}\right)  \mathrm{U}_{2j_{1}}\left(  \frac
{\sigma_{2}}{2}\right) \\
&  \ \exp\left\{  -\left(  \sigma_{k}-\left\langle \chi\right\rangle \right)
\alpha_{k,r}\left(  \sigma_{r}-\left\langle \chi\right\rangle \right)
\right\}  d^{2}\sigma,\nonumber
\end{align}
where matrix $\alpha_{k,r}$ is given by
\begin{equation}
\left(
\begin{array}
[c]{cc}%
\alpha_{1,1} & \alpha_{1,2}\\
\alpha_{2,1} & \alpha_{2,2}%
\end{array}
\right)  =\frac{1}{\left(  A_{0,0}^{-1}\right)  ^{2}-\left(  A_{0,R}%
^{-1}\right)  ^{2}}\left(
\begin{array}
[c]{cc}%
A_{0,0}^{-1} & -A_{0,R}^{-1}\\
-A_{0,R}^{-1} & A_{0,0}^{-1}%
\end{array}
\right)  ,
\end{equation}
one may do computation expanding over powers $A_{0,R}^{-1}$ and taking into
account simple relation $\chi^{\left(  j\right)  }\chi=\chi^{\left(
j+\frac{1}{2}\right)  }+\chi^{\left(  j-\frac{1}{2}\right)  }$. In particular
for magnetization $M_{j,j^{\prime}}$ we get
\begin{equation}
M_{j,j^{\prime}}\equiv\left\langle \chi_{0}^{\left(  j\right)  }\chi_{\infty
}^{\left(  j^{\prime}\right)  }\right\rangle =\left\langle \chi^{\left(
j\right)  }\right\rangle \left\langle \chi^{\left(  j^{\prime}\right)
}\right\rangle .
\end{equation}

In the confinement area $\left(  \gamma<\gamma_{c}\right)  $ for integer
$j_{1,2}$ $=\left[  j_{1,2}\right]  $ for $\alpha R\sim1$%

\begin{equation}
\ \left\langle \chi_{1}^{\left(  j_{1}\right)  }\chi_{2}^{\left(
j_{2}\right)  }\right\rangle \ \simeq\left\langle \chi^{\left(  j_{1}\right)
}\right\rangle \left\langle \chi^{\left(  j_{2}\right)  }\right\rangle
\exp\left\{  \left(  j_{1}j_{2}-\frac{1}{4}\right)  \frac{2\left(
A_{0,R}^{-1}\right)  ^{2}}{\left(  8-\frac{1}{\gamma}\right)  ^{2}}\right\}
\ \ . \label{casi1}%
\end{equation}

In particular for $j_{1}=j_{2}$ $=j$ we obtain for $\alpha R\sim1$
\begin{equation}
\ \left\langle \chi_{1}^{\left(  j\right)  }\chi_{2}^{\left(  j\right)
}\right\rangle \ \simeq\left\langle \chi^{\left(  j\right)  }\right\rangle
^{2}\exp\left\{  \frac{2\left(  A_{0,R}^{-1}\right)  ^{2}}{\left(  8-\frac
{1}{\gamma}\right)  ^{2}}C_{2}\right\}  \sim\exp\left\{  -\frac{\alpha RC_{2}%
}{\left(  4-\frac{1}{2\gamma}\right)  ^{2}}\right\}  \ \ \label{casi2}%
\end{equation}
where $C_{2}=\left(  j+\frac{1}{2}\right)  ^{2}-\frac{1}{4}$ is quadratic
Casimir operator.

For half-integer $j_{1,2}=$ $\left[  j_{1,2}\right]  +\frac12$%

\begin{equation}
\ \ \left\langle \chi_{x_{1}}^{\left(  j_{1}\right)  }\chi_{x_{2}}^{\left(
j_{2}\right)  }\right\rangle \simeq\frac{2\left\langle \chi\chi^{\left(
j_{1}\right)  }\right\rangle \left\langle \chi^{\left(  j_{2}\right)  }%
\chi\right\rangle }{\left(  A_{0,0}^{-1}\right)  ^{2}}A_{0,R}^{-1}=q\left(
j_{1};j_{2}\right)  A_{0,R}^{-1},
\end{equation}
with
\begin{equation}
q\left(  j_{1};j_{2}\right)  =\frac{2\left(  \left\langle \chi^{\left(
j_{1}+\frac{1}{2}\right)  }\right\rangle +\left\langle \chi^{\left(
j_{1}-\frac{1}{2}\right)  }\right\rangle \right)  \left(  \left\langle
\chi^{\left(  j_{2}+\frac{1}{2}\right)  }\right\rangle +\left\langle
\chi^{\left(  j_{2}-\frac{1}{2}\right)  }\right\rangle \right)  }{\left(
8-\frac{1}{\gamma}\right)  ^{2}},
\end{equation}
and
\begin{equation}
\left\langle \chi^{\left(  j-\frac{1}{2}\right)  }\right\rangle \simeq\frac
{3}{4}\frac{\left(  2j-1\right)  !}{\left(  j-\frac{1}{2}\right)  !}\left(
2\left(  1-\frac{1}{8\gamma}\right)  \right)  ^{j-\frac{1}{2}}.
\end{equation}

Therefore taking into account
\begin{equation}
\left\langle \chi^{\left(  j\right)  }\chi\right\rangle =\left\langle
\chi^{\left(  j+\frac{1}{2}\right)  }\right\rangle +\left\langle \chi^{\left(
j-\frac{1}{2}\right)  }\right\rangle \simeq\left\langle \chi\right\rangle
\ ^{2j+1}+\left\langle \chi\right\rangle \ ^{2j-1},
\end{equation}
we obtain%

\begin{align}
\ \ \left\langle \chi_{x_{1}}^{\left(  j_{1}\right)  }\chi_{x_{2}}^{\left(
j_{2}\right)  }\right\rangle  &  \simeq\ \ \left\langle \chi^{\left(
j_{1}\right)  }\right\rangle \ \ \left\langle \chi^{\left(  j_{2}\right)
}\right\rangle +\frac{2\left\langle \chi\chi^{\left(  j_{1}\right)
}\right\rangle \left\langle \chi^{\left(  j_{2}\right)  }\chi\right\rangle
}{\left(  A_{0,0}^{-1}\right)  ^{2}}A_{0,R}^{-1}\\
&  \simeq\ \ \left\langle \chi\right\rangle \ ^{2j_{1}+2j_{2}}\ \exp\left\{
8\gamma^{2}\left(  \left\langle \chi\right\rangle +\frac{1}{\left\langle
\chi\right\rangle }\right)  ^{2}A_{0,R}^{-1}\right\}  .\nonumber
\end{align}
In particular we get
\begin{equation}
\left\langle \chi^{\left(  1\right)  }\right\rangle +1=\left\langle \chi
^{2}\right\rangle =\left[  \left(  \frac{\eta_{x}A_{x,0}^{-1}}{2}\right)
^{2}\right]  _{\eta_{x}=\eta_{0}\rightarrow0}+\frac{1}{2}A_{0,0}%
^{-1}=\left\langle \chi\right\rangle ^{2}+\frac{1}{2}A_{0,0}^{-1},
\end{equation}
and
\begin{align}
\left\langle \chi_{0}^{2}\chi_{R}^{2}\right\rangle  &  =\left[  e^{-\frac
{1}{4}\eta A^{-1}\eta}\frac{\partial^{2}}{\partial\eta_{R}^{2}}\frac
{\partial^{2}}{\partial\eta_{0}^{2}}e^{\frac{1}{4}\eta A^{-1}\eta}\right]
_{\eta_{x}=\eta_{0}\rightarrow0}\\
&  =\left(  \left\langle \chi\right\rangle ^{2}+\frac{1}{2}A_{0,0}%
^{-1}\right)  ^{2}+\frac{1}{2}\left(  A_{R,0}^{-1}\right)  ^{2}+\frac{3}%
{2}A_{0,R}^{-1}\left\langle \chi\right\rangle ^{2},\nonumber
\end{align}
therefore, correlation function for characters in the adjoint representation
$\left\langle \chi_{0}^{\left(  1\right)  }\chi_{R}^{\left(  1\right)
}\right\rangle $ may be written as
\begin{equation}
\left\langle \chi_{0}^{\left(  1\right)  }\chi_{R}^{\left(  1\right)
}\right\rangle \equiv\left\langle \chi_{0}^{2}\chi_{R}^{2}\right\rangle
-2\left\langle \chi^{2}\right\rangle +1=\left\langle \chi^{\left(  1\right)
}\right\rangle ^{2}+\frac{1}{2}\left(  A_{R,0}^{-1}\right)  ^{2}+\frac{3}%
{2}A_{0,R}^{-1}\left\langle \chi\right\rangle ^{2}.
\end{equation}

\end{document}